\documentclass[12pt]{article}
\usepackage{amsmath}
\usepackage{amsfonts}
\usepackage{amssymb, amsthm}
\usepackage{graphicx,epsfig}

\def\max{\mathrm{max}}
\def\min{\mathrm{min}}
\def\e{{\bf{e}}}
\def\n{{\bf{n}}}
\def\b{{\bf{b}}}
\def\t{{\bf{t}}}
\def\j{{\bf{j}}}

\begin{document}
\title{Role of the mean curvature in the geometry of magnetic confinement configurations}
\author{A.A. Skovoroda \thanks{Russian Research Centre Kurchatov Institute, pl. Kurchatova 1,
Moscow, 123182 Russia; e-mail: skovorod@nfi.kiae.ru} \and I.A. Taimanov \thanks{Sobolev Institute
of Mathematics, Novosibirsk, 630090 Russia; e-mail: taimanov@math.nsc.ru}}
\date{}
\maketitle

\abstract{Examples are presented of how the geometric notion of the
mean curvature is used for general magnetic field configurations and
magnetic surfaces. It is shown that the mean magnetic curvature is
related to the variation of the absolute value of the magnetic field
along its lines. Magnetic surfaces of constant mean curvature are
optimum for plasma confinement in multimirror open confinement
systems and rippled tori.}

\section*{Introduction}

The mean curvature is one of the basic notions in the geometry of
surfaces and vector fields [1, 2], while the geometry of magnetic
fields, as applied to the problems of fusion plasma confinement,
does not in fact invoke this notion (as is confirmed in [3]). The
object of the present methodological note is to bridge this gap. We
begin by giving the definitions and main results on the differential
geometry of surfaces in three-dimensional Euclidean space and of
vector fields and then present various applications of the notion of
mean curvature in magnetic confinement systems.

\section{Mean curvature of surfaces and vector fields}

 The mean
curvature of a surface, $H_S$, is a local quantity defined by the
formula
$$
H_S = -\frac{1}{2}(k_\min + k_\max),
\eqno{(1)}
$$
where $k_\min$ and $k_\max$ are the minimum and maximum
curvatures of the lines of intersection of the surface with mutually
perpendicular planes passing through the normal to the surface at
its given point. For $\e \cdot (\nabla \times \e) =0$, draw a family of
surface orthogonal to a unit vector field $\e$ (such that $|\e| = 1$).
The mean curvature of these surfaces is called the mean curvature of
the vector field $\e$. In the general case $\e \cdot (\nabla \times \e) \neq 0$
and the mean curvature $H$ of a vector field $\e$ is defined by the formula
[2]
$$
H = -\frac{1}{2} \nabla \cdot \e.
\eqno{(2)}
$$
For a unit vector $\e = \n$ along the normal $\n$ to a surface, the
general definition of the curvature $H$ coincides with that of $H_S$
(see the definition (1)). The surface of zero mean curvature, $H_S =
0$, is called a minimal surface. Examples of minimal surfaces are
given by surfaces of minimum area $S$ with a fixed boundary (soap
films). Closed minimal surfaces do not exist. Among surfaces of
constant mean curvature $H_S = \mathrm{const}$ are soap films
between media at different pressures $p$, in which case the mean
curvature is just the pressure difference. Another example of the
surfaces of constant mean curvature $H_S = \mathrm{const}$ are those
among the surfaces bounding regions of given volume $V$ that have a
minimum area $S$. Such surfaces are called isoperimetric profiles
and are perfect spheres of constant curvature. The Aleksandrov
theorem [4] states that such spheres are the only embedded (or
nested, i.e. non-self-intersecting) closed surfaces of constant mean
curvature. A consequence of the theorem is, in particular, that
embedded (nested) tori of constant mean curvature do not exist.

The statement of this general theorem can be
simply verified for surfaces of revolution. Assuming that the axis
of revolution is the $Z$ axis of a cylindrical coordinate system, we
specify the shape of a surface of revolution by the equation $\Phi=
z-z(r)$. Substituting the normal to this surface, $\n = \nabla\Phi/|\nabla\Phi|$, into
(2), we obtain the equation
$$
2H_S = \frac{1}{r}\frac{d}{dr} \left(\frac{rz^\prime}{\sqrt{1+z^{\prime 2}}}\right),
\eqno{(3)}
$$
where the prime
denotes the derivative with respect to the radius $r$. For a constant
mean curvature, Eq. (3) is integrable:
$$
H_Sr^2 + C = \frac{rz^\prime}{\sqrt{1+z^{\prime 2}}},
\eqno{(4)}
$$
where $C$ is a constant
of integration. Equations (3) and (4) imply that a closed plane
curve having two or more points where $z^\prime = 0$ at $r > 0$ does not
exist. Consequently, a torus of constant mean curvature is
impossible. Unfortunately, in considering an axisymmetric example in
[5], sad mistakes were made that led to an erroneous conclusion
about the existence a torus of constant mean curvature.

Equation (4)
shows that there may be two points where $z^\prime = \infty$
at $r > 0$. The generating contours of these surfaces of revolution
are described by a focus of a hyperbola
\footnote{A rolling of a
hyperbola determines one period of continuous curve in Fig. 1.}
or an ellipse (see Fig. 1) rolled along the straight axis of
revolution [6].

\begin{figure}[ht]
\begin{center}
\epsfig{file=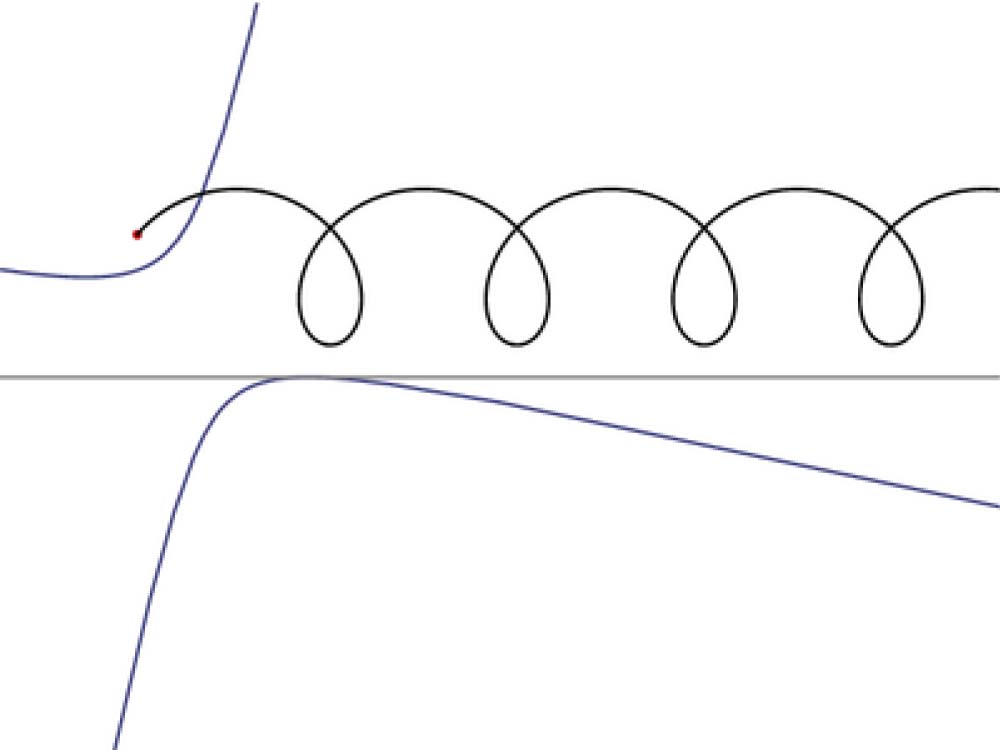,height=30mm,clip=} \hskip1cm
\epsfig{file=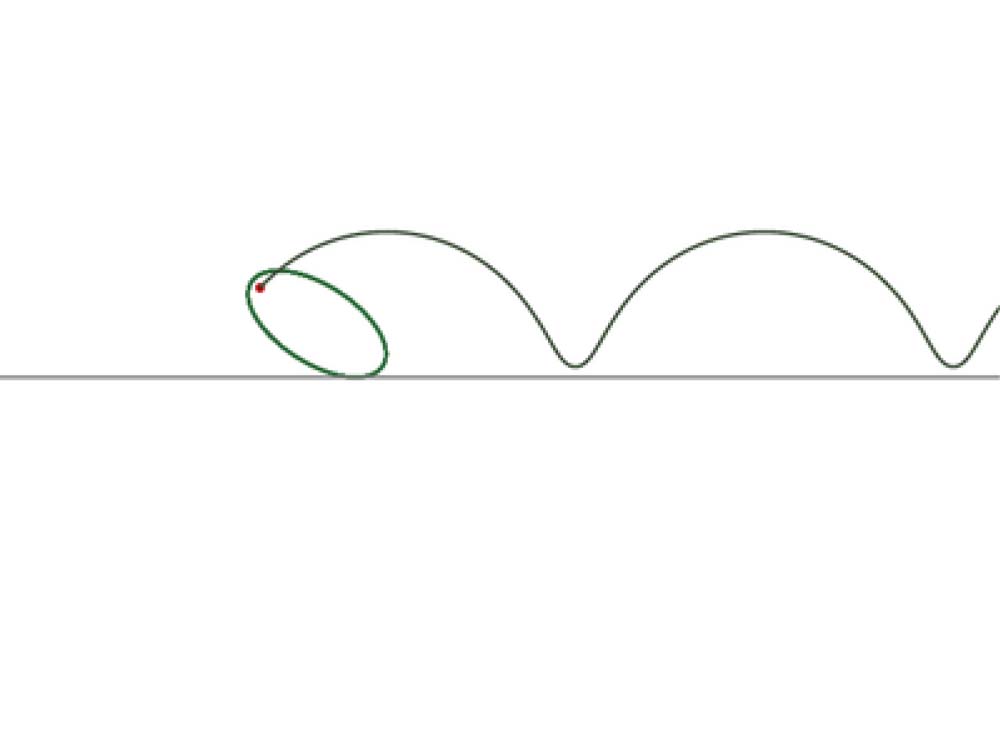,height=30mm,clip=} \caption{Rolling of
hyperbola and ellipse.}
\end{center}
\end{figure}

\noindent A cylinder is a limiting case of rolling of a circle. This
demonstrates the existence of surfaces with $H_S = \mathrm{const}$
that are periodic along the axis of revolution (see Fig. 2).

\begin{figure}[ht]
\begin{center}
\epsfig{file=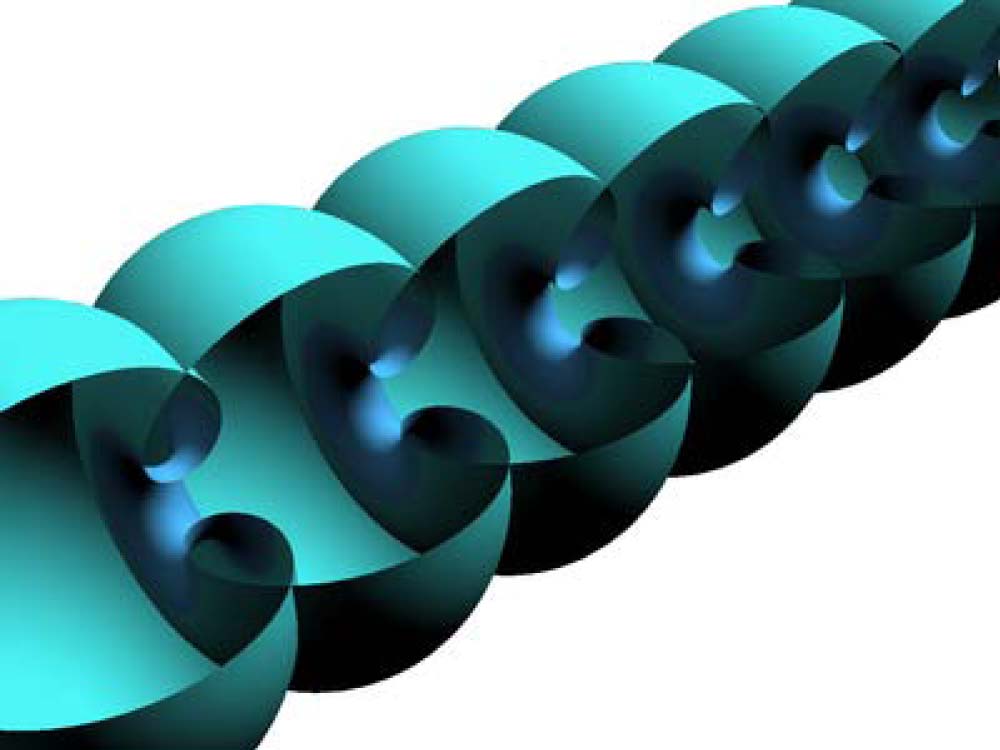,height=35mm,clip=} \hskip1cm
\epsfig{file=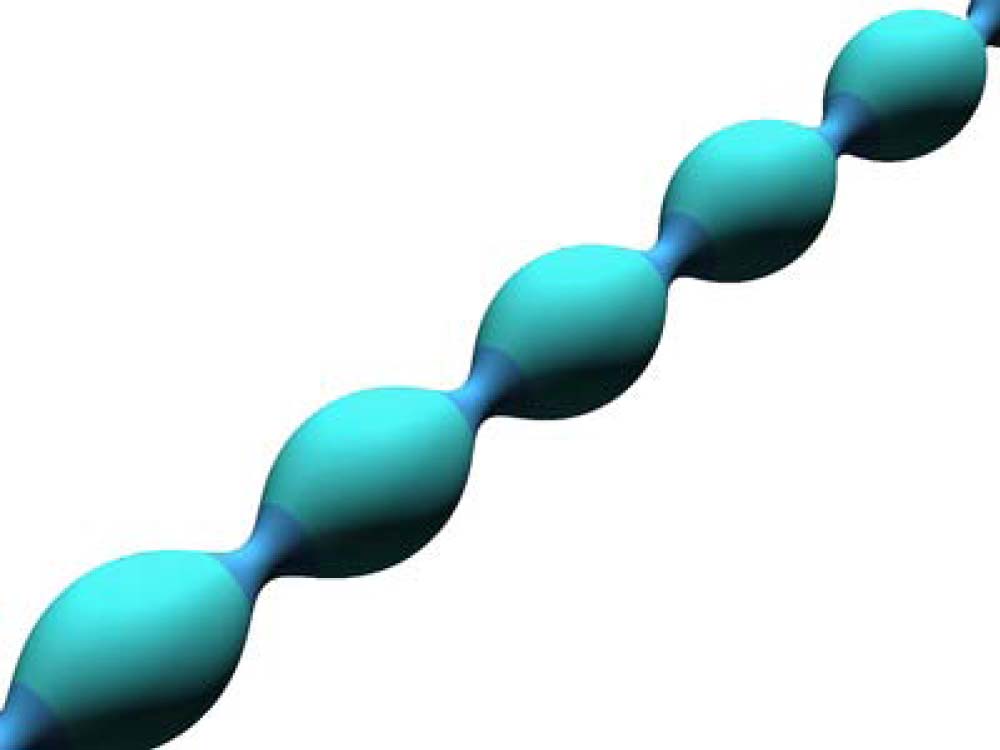,height=35mm,clip=} \caption{Nodoid and
unduloid.}
\end{center}
\end{figure}

In the theory of surfaces, the Willmore functional determined by an
integral of the square of the mean curvature over a surface plays an
important role. For closed surfaces, this functional is conformally
invariant: in a conformal mapping of a three-dimensional space onto
itself, the values of the Willmore functional on a surface and on
its transform coincide [7]. This functional not only plays a key
role in the Weierstrass representation of surfaces [7] but has
recently come into use in biophysics and colloidal
chemistry---disciplines in which it is known as the Helfrich
functional [8]. The critical points of the Helfrich functional
generate Willmore surfaces [7]. In contrast to minimal surfaces,
there are many examples of closed Willmore surfaces, including
embedded (non-self-intersecting) ones. Among them are, e.g., all
spheres of constant curvature and also a Clifford torus --- a
surface of revolution generated by revolving such circle about its
axis that the ratio of the distance $R$ between the center of the
circle and the axis of revolution to the circle radius $r$ is
$\sqrt{2}$.

\section{Mean curvature of a magnetic field}

According to (2), a magnetic field ${\bf B} = B\cdot \b$, where $|\b| = 1$
and $B=|{\bf B}|$ is the absolute value of fiel, can be characterized by
the mean curvature
$$
H_b = \b \cdot \nabla \log \sqrt{B}.
\eqno{(5)}
$$
The variation of the absolute value of the
magnetic field along its lines is an important parameter of plasma
magnetic confinement systems and is often used in the geometry of
magnetic fields [3, 9]. For a vacuum magnetic field ${\bf B} = -\nabla \varphi$, where
$\varphi$ is the scalar magnetic potential, the mean curvature $H_b$
coincides with that of an equipotential surface $\varphi = \mathrm{const}$ for
which the vector $\b$ is a unit normal vector. Note that, by (2), the expression (5) for the mean
curvature is also valid for $\nabla \times {\bf B} \neq 0$.

At the extreme points
of the absolute value of the magnetic field along its lines, we have
$H_b = 0$. In the so-called isodynamic toroidal configurations
revealed by D. Palumbo [10], the absolute value of the magnetic
field is constant on the nested magnetic surfaces, $B = \mathrm{const}$, so
the mean curvature is zero: $H_b = 0$, over the entire confinement
region. Such isodynamic toroidal configurations are possible only in
the presence of the discharge current [11]. Without the discharge
current, i.e., in vacuum, the equipotential surfaces in an
isodynamic configuration should be minimal, which is impossible,
however, in view of the results obtained by Palumbo [11].

 The
magnetic fields that form a family of nested magnetic surfaces,
${\bf B} \cdot \n = 0$, in a finite spatial region play a governing role in
plasma confinement. Magnetic configurations for plasma confinement
can be divided into open configurations with rippled cylindrical
nested magnetic surfaces and closed configurations with nested
toroidal magnetic surfaces of complicated shape. The solenoidal
nature of the magnetic field implies that the toroidal magnetic flux
$\Phi$ within a magnetic surface is conserved. This is why the most
general equation for a family of nested magnetic surfaces is
formulated in terms of the toroidal magnetic flux: $\Phi(x,y,z)=\mathrm{const}$.
The function $\Phi$ is a single-valued solution, if there is
any, to the equation ${\bf B} \cdot \nabla \Phi = 0$ with a known
magnetic field having a nonzero rotational transform. Hence, at each
point of the plasma confinement region, the vector field of unit
vectors normal to the magnetic surfaces, $\n = \nabla \Phi/|\nabla \Phi|$ such that
$\n \cdot (\nabla \times \n)=0$, is usually defined. By substituting the vector
$\n$ into (2), it is possible to determine the mean curvature
$H_S$ of the magnetic surfaces---a quantity that plays an important
role in the theory of plasma confinement systems.

For completeness
sake, we supplement the vectors $\b$ and $\n$, which have been
introduced above, with the binormal vector $\t = \b \times \n$ up to an
orthonormalized magnetic basis. The vector $\t$ is directed along the
vector ${\bf B} \times \nabla \Phi$, which is orthogonal to ${\bf B}$.
In an equilibrium state in which the plasma currents $\j$
flow along the magnetic surfaces $\j \cdot \nabla \Phi=0$ the
vector ${\bf B} \times \nabla \Phi$ is solenoidal and its lines lie on the
magnetic surfaces [9]. Applying (2) to $\t$
defines the mean curvature of the additional field:
$$
H_t = \t \cdot \nabla \log \sqrt{B|\nabla \Phi|}.
\eqno{(6)}
$$
This curvature is related to such familiar parameter as the geodesic
curvature of the magnetic field lines [9].

\subsection{Magnetic surfaces with $H = \mathrm{const}$}

 The energy
and particle confinement in magnetic systems is commonly
characterized by integral confinement times. For definiteness, let
us consider the particle confinement time $\tau_N$. This time is
calculated from the formula $\tau_N = N/I$, where $N$ is the total
number of particles in the system after the injection of the
particle current $I$. In turn, the total number of particles is $N =
nV$, where $n$ is the mean particle density in the system and $V$ is its
volume. Since, in a steady state, the injection current is equal to
the loss current, we have $I = nvS$, where $v$ is the mean velocity with
which the particles escape from the system through a boundary region
of area $S$. As a result, the particle confinement time is given by
the formula $\tau_N = V/vS$. The better the confinement, the longer
this time. Under the assumption that $v$ is constant, we arrive at
the conclusion that the boundary surfaces of constant mean curvature
are optimum for plasma confinement.

As shown in Section 1, there
exist periodic axisymmetric unduloid surfaces with $H_S = \mathrm{const}$ (see
Fig. 2) that can be used to optimize plasma confinement in ambipolar
open magnetic systems [9]. The central solenoid of such
configurations with a straight magnetic axis is characterized by the
length $L$ and on-axis mirror ratio $P = B_\max/B_\min$, or the ratio of the
maximum and minimum radii $r_\max/r_\min = \sqrt{P}$.
The values of $L$ and $P$
determine the parameters of an ellipse that generates an optimum
unduloid for a given geometry of the magnetic configuration.

In Section 1, it was pointed out that tori with $H_S = \mathrm{const}$ do not
exist. It is, however, for an ambipolar open magnetic confinement
system that we consider asymptotic surfaces of constant mean
curvature. Figure 3 shows a way how to close open systems by means
of the magnetic surface of a Kadomtsev rippled confinement system
[12].

\begin{figure}[ht]
\begin{center}
\epsfig{file=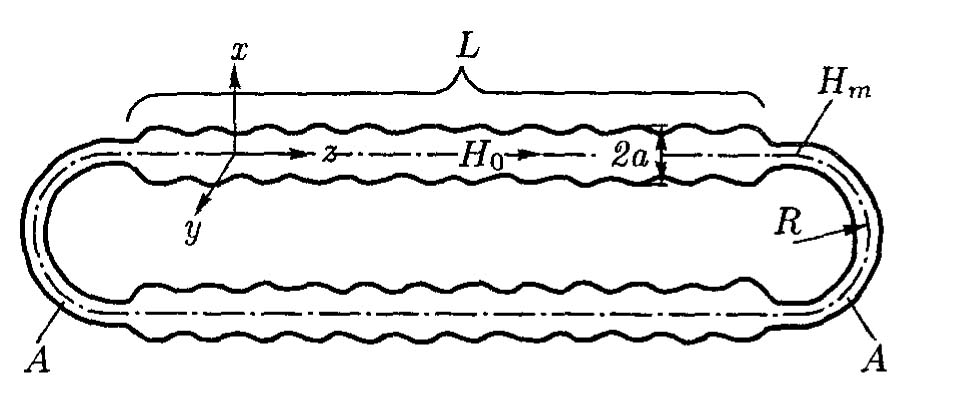,height=35mm,clip=} \caption{The Kadomtsev
confinement system.}
\end{center}
\end{figure}

\noindent
Let us modify the Kadomtsev system as follows. On two long
straight portions, it is possible to use unduloids with a relatively
weak mean magnetic field, and curvilinear elements can be half-tori
with a magnetic field strong enough for the radius of the tori to be
small. The system can also be arranged to have a square shape: four
straight portions with unduloids are closed by curvilinear elements
in the form of quarter-tori. This method of modifying the Kadomtsev
system can be repeated over and over again. The main idea is that,
for a sufficiently large ratio $L/L_c$, where $L_c$ is the length of the
curvilinear elements, and a sufficiently large mirror ratio $P = B_c/B$,
where $B_c$ is the magnetic field in a curvilinear element, the
ratio $V/S$ is asymptotically determined only by the straight
portions of the system.

\subsection{Calculation of the shape of the magnetic
Surfaces}

 The main plasma confinement problem is to calculate a
distortion of the shape of the magnetic surfaces that are in
equilibrium with a plasma of isotropic pressure, $p = p(\Phi)$. The
pressure is assumed to be constant on the magnetic surfaces because
it is equalized by a mechanism associated with the fast motion of
charged plasma particles along magnetic field lines with irrational
rotational transforms. An increase in pressure distorts the magnetic
surfaces and, as a rule, destroys their nested structure, thereby
deteriorating the plasma confinement radically. This is why solving
such problems is aimed at determining the maximum volume-averaged
value of the parameter $\beta_{\mathrm{eq}} \sim \left\langle
\frac{p}{B^2}\right\rangle$ that is consistent with
equilibrium. This maximum beta value, and consequently, the plasma
confinement efficiency, is very sensitive to the geometric shape of
the magnetic surfaces. To be specific, we can mention innovative
stellarators with a complicated three-dimensional geometry of
helical tori---devices in which the maximum $\beta_{\mathrm{eq}}$ values are
more than an order of magnitude higher than those in tokamaks (see
[9] for comparison). That is why it is important to search for
special geometries of magnetic confinement systems. In this way, it
may be very helpful to use Eq. (2) written for the normal to a
magnetic surface in the form of an elliptic equation for determining
$\Phi(x,y,z)$ at given $|\nabla\Phi|$ and $H_S$:
$$
\nabla \cdot \left(\frac{\nabla\Phi}{|\nabla\Phi|}\right) = - 2H_S.
\eqno{(7)}
$$

Let us
introduce the function $\delta = \Phi/|\nabla\Phi|$ which is in fact the distance
between neighboring magnetic surfaces. The function $\delta$ so
introduced is required to calculate the magnetic field ${\bf B}$ on a given
magnetic surface [13]. Instead of $\delta$, it is possible to use
the absolute value of the magnetic field $B$ [14]. This is why the
choice of the function $|\nabla\Phi|$ is largely determined by how
the absolute value of the magnetic field varies along its lines.
Moreover, the pseudosymmetry condition, which is now widely used to
synthesize innovative stellarators, is also constructed based on the
geometry of isomagnetic surfaces (i.e., the surfaces of constant
magnetic field strength $B$) [9].

The function $H_S$ determines the  geometry of the confinement system.
From an expression obtained in
[14], specifically,
$$
H_S = \frac{\kappa_n + \hat{\kappa}_n}{2},
\eqno{(8)}
$$
where $\kappa_n$ and $\hat{\kappa}_n$
are the normal curvatures of the lines of the magnetic field ${\bf B}$ and
its complement ${\bf B} \times \nabla \Phi$, it is clear that the pressure $p$
enters Eq. (7) through the mean curvature. In fact, the
(equilibrium) force balance equation yields the equalities
$$
\kappa_n = \frac{1}{B^2} \n
\cdot \nabla\left(p+\frac{B^2}{2}\right) =
\frac{|\nabla\Phi|}{B^2}\, \frac{dp}{d\Phi} + \n
\cdot \nabla\log B.
\eqno{(9)}
$$
The expression for $\hat{\kappa}_n = \n \cdot (\t \cdot \nabla)\t$
takes the simplest form for
axisymmetric currentless configurations [15]:
$$
\hat{\kappa}_n = -\ \cdot \nabla \log \frac{|\nabla\Phi|}{B} =
-\n \cdot \nabla\log \frac{|\nabla\Phi|}{B^2} - \n\cdot \nabla \log B.
\eqno{(10)}
$$
Simple
manipulations put Eq. (7) into the form
$$
\nabla \cdot \frac{\nabla\Phi}{(|\nabla\Phi|^2/B^2)} = -\frac{dp}{d\Phi}.
\eqno{(11)}
$$
Since, in
cylindrical coordinates, we have $|\nabla\Phi|^2/B^2 = r^2$ up to constants, we
arrive at the Grad--Shafranov equation.

Hence, after some algebraic
manipulations, Eq. (7) can be used not only to search for an optimum
geometry of the boundary magnetic surface but also to solve
equilibrium problems.

\subsection{The integral $\oint_S H_S dS$ on nested magnetic surfaces}

Let us consider the radial variation of the integral $\oint_S H_S dS$. To do this,
we use the divergent expression presented in [2] for the Gaussian
curvature $K$ of a unit vector field $\e$:
$$
K = \frac{1}{2}\nabla \cdot (2H\e +\kappa).
\eqno{(12)}
$$
where $\kappa = -\e \times(\nabla \times \e)$
is the curvature vector of the lines of the vector field $\e$. Setting
$\e = \n$, where $\n$ is the normal to a family of nested magnetic
surfaces $\Phi$, and integrating the curvature from formula (12)
over the volume $dV = \frac{dS\,d\Phi}{|\nabla\Phi|}$
between two neighboring magnetic surfaces
$\Phi$ and $\Phi+d\Phi$, we obtain the equalities
$$
\oint_V KdV = d\Phi \oint_S \frac{K}{|\nabla \Phi|} dS = d\oint_S H_S dS.
\eqno{(13)}
$$
Here, we
have used the fact that the magnetic flux between the surfaces is
constant, $d\Phi = \mathrm{const}$. The result is
$$
\frac{d\oint_S H_SdS}{d\Phi} = \oint_S\frac{K}{|\nabla\Phi|}dS.
\eqno{(14)}
$$
Since $\oint_S KdS =0$ for
toroidal surfaces, and since $|\nabla \Phi| \approx \mathrm{const}$,
the integral on the right-hand side of formula (14) vanishes.

\section*{Conclusions}

The mean curvature of the magnetic field vector is related to the
variation of the absolute value of the magnetic field along its
lines. In the presence of magnetic surfaces and, consequently, of
the orthonormalized magnetic basis ($\b, \n, \t$), the mean curvature
can be introduced for each basis vector. The mean curvature of the
normal vector coincides with that of the magnetic surface. Magnetic
surfaces of constant mean curvature, having a minimum surface area
at a fixed volume, are optimum for plasma confinement in multimirror
open systems and rippled tori with straight portions. By specifying
the mean curvature of the magnetic surfaces and the distance to the
nearest magnetic surface, it is possible to calculate the shape of
the magnetic surfaces. All this goes to show that it may be helpful
to use the notion of the mean curvature in the geometry of magnetic
fields in plasma magnetic confinement systems.

\vskip7mm

{\bf Acknowledgments.} We are grateful to N. Schmitt for permission
to borrow Figs. 1 and 2. This work was supported in part by the
Russian Foundation for Basic Research, the Federal Special-Purpose
Program ``Scientific and Pedagogical Personnel of the Innovative
Russia for 2009--2012'', Presidium of the Russian Academy of
Sciences (under the program "Fundamental Problems of Nonlinear
Dynamics"), and the Council of the Russian Federation Presidential
Grants for State Support of Leading Scientific Schools (project no.
NSh-65382.2010.2).

\end{document}